\documentclass[11pt]{article}

\usepackage[latin1]{inputenc}
\usepackage{epsfig}
\usepackage{amsfonts}
\usepackage{amssymb}
\usepackage{subfigure}
\usepackage[usenames]{color}
\usepackage{soul}
\usepackage{comment}
\usepackage{multirow}
\usepackage[cmex10]{amsmath}
\usepackage{bm}
\usepackage{setspace}
\usepackage{fullpage}
\usepackage{amsmath}
\usepackage{wrapfig}

\newcommand{\Pmat}{\mathbf{P}}

\newcommand{\pr}[1]{\Pr \left\{{#1}\right\}}

\def\piD{\boldsymbol{\pi}^d}
\def\piS{\boldsymbol{\pi}^s}

\def\bigghost{\text{\white{\small{!}}}}

\newcommand{\white}[1]{{\color{white} #1}}

\newcommand{\per}[1]{$p_{\text{e}}$}

\newcommand{\stirlingtwo}[2]{\genfrac{\{}{\}}{0pt}{}{#1}{#2}}

\begin{document}


\title{Performance Analysis of a Multiuser Multi-Packet Transmission System for WLANs in Non-Saturation Conditions}

\date{}


\author{Boris Bellalta$^{(1)}$, Azadeh Faridi, Jaume Barcelo, Vanesa Daza, Miquel Oliver\\
Department of Information and Communication Technologies. \\Universitat Pompeu Fabra, Barcelona, Spain\\(1) Corresponding author, e-mail: boris.bellalta@upf.edu }

\maketitle

\begin{abstract}
Multiuser Multi-Packet Transmission (MPT) from an Access Point (AP) equipped with multiple antennas to multiple single-antenna nodes can be achieved by exploiting the spatial dimension of the channel. In this paper we present a queueing model to analytically study such systems from the link-layer perspective, in presence of random packet arrivals, heterogeneous channel conditions and packet errors. The analysis relies on a blind estimation of the number of different destinations among the packets waiting in the queue, which allows for building a simple, but general model for MPT systems with per-node First-In First-Out (FIFO) packet scheduling. Simulation results validate the accuracy of the analytical model and provide further insights on the cross-relations between the channel state, the number of antennas, and the number of active users, as well as how they affect the system performance. The simplicity and accuracy of the model makes it suitable for the evaluation of Medium Access Control (MAC) protocols for Ad-Hoc or Wireless Local Area Networks supporting multiuser MPT in non-saturation conditions, where the queueing dynamics play an important role on the achieved performance, and simple user selection algorithms are required.


\end{abstract}



\onehalfspacing

%
%

\section{Introduction} \label{Sec:Intro}

In packet-based wireless networks, the use of spatial multiplexing allows for simultaneous transmission of multiple packets, directed to a single or multiple destinations. In this paper, we focus on a scenario where an AP equipped with multiple antennas is able to simultaneously transmit multiple packets, each directed to a different single-antenna node. This scenario is known as Multiuser Multi-Packet Transmission (MPT), a packet-based extension of Downlink Space Division Multiple Access (DL-SDMA).

Research on Multiuser MPT has mainly been focused on the design of efficient joint precoding and user selection strategies, as a trade-off between computational complexity and the ability to maximize the system capacity, i.e., the number of bits per Hertz of available bandwidth that can be successfully transmitted over the channel. {A comprehensive survey of such works has been presented in  \cite{mietzner2009multiple}}. In these works, it is usually assumed that the transmitter has separate per-user queues that are always saturated. This reduces the problem to finding the set of users that maximize the system sum-rate capacity, based on the current state of the channel. Specific schemes for user selection range from random selection to greedy schedulers that benefit from the existing multiuser diversity \cite{gesbert2007shifting}.

The main advantage of random schedulers is their simplicity, as the specific channel conditions of the destinations are not considered for selecting the set of destinations at each transmission. Therefore, they do not need to have recent channel information from all potential destinations, but only from the ones selected for transmission, reducing the overhead required to obtain and keep this information updated. In addition, it provides a fair channel access to the competing users, as all users are selected only based on their traffic load, regardless of their instantaneous channel conditions. For these reasons, they are specially suitable for Ad-Hoc Networks or WLANs, where on the one hand, fairness and simplicity are design requirements and, on the other hand, the sporadic and bursty traffic patterns may not allow for taking full advantage of the existence of Channel State Information (CSI) at the transmitter from all potential destinations. This approach has been widely considered in the design of MAC protocols for WLANs supporting MPT by extending the RTS/CTS mechanism \cite{cai2008distributed,li2010multi}.

There are still very few works that address spatial multiplexing from the link-layer perspective, and even fewer that focus on the queueing dynamics. In \cite{zhou2008queuing}, the authors present a point-to-point MIMO system, considering both transmit diversity (STBC) and spatial multiplexing (BLAST) schemes for a single-user MPT. A similar work that overcomes the approximation done in \cite{zhou2008queuing} for the BLAST scenario is presented in \cite{zhao2012performance}, although the presented results are only valid for two antennas. To the best of our knowledge, \cite{rashid2009cross} is the only work where a detailed queueing model for Multiuser MPT systems is presented. In \cite{rashid2009cross}, the nodes are assumed to be equipped with at least as many antennas as the AP. Therefore, the AP can use its multiple antennas to send multiple packets to a single or multiple users simultaneously, following the independent stream scheduler approach \cite{heath2001multiuser}. Using the independent stream scheduler, destinations are selected based on the CSI, allowing the transmission of multiple packets to the same destination if in that way the throughput is maximized.

In \cite{bellalta2009space}, an M/G/1 batch-service queueing model was proposed to characterize the behavior of MPT systems. However, the queueing model assumes that always the maximum number of packets can be transmitted, regardless of the number of destinations represented among the buffered packets at the AP. Therefore, the model is only valid for when the nodes are equipped with at least as many antennas as the AP. The same model was then used in \cite{bellalta2010upper,bellalta2012role} to study the queueing behavior in multiuser MPT systems when all the receivers have a single antenna. This implies that each transmission can contain at most one packet per destination. Comparing with the analytical results from \cite{bellalta2009space}, the simulation results in \cite{bellalta2010upper} and \cite{bellalta2012role} show the loss in performance due to having a single antenna at each destination. In \cite{bellaltaapproximate}, this performance loss is modeled analytically, by using an estimate of the number of different destinations represented among the packets waiting in the queue at the AP, which determines the number of packets that can be scheduled in each transmission. This approach was further validated in \cite{bellalta2011buffer}, where the effect of the buffer size in such systems, as well as in the accuracy of the analytical model presented, is evaluated. In all those works an ideal channel is considered, and therefore, the negative effect, in terms of lower transmission rates, caused by the simultaneous transmission of multiple packets is not included in the analysis.

In this paper, we extend the model and the results presented in \cite{bellaltaapproximate,bellalta2011buffer} by providing a detailed analytical model that includes a realistic channel model, supports heterogeneous channel conditions, multiple transmission rates, packet errors, and a tunable scheduler based on the observed channel conditions in a scenario that consists of a multiple-antenna AP and single-antenna users. Moreover, new insights on the model accuracy in terms of the number of nodes and buffer size are provided, considering heterogeneous traffic and nodes with heterogeneous channel conditions. Hence, the presented model can be used to understand and evaluate the different interactions that exist in a Multiuser MPT system between the traffic load, the buffer size, the number of antennas at the AP, the number of different nodes, the channel characteristics, and the protocol overheads required for CSI estimation and reporting. In addition, due to the model characteristics, it can be easily coupled with other link-layer mechanisms to evaluate more complex systems.

The paper is structured as follows. The scenario, together with the system model and the assumptions considered, is introduced in Section \ref{Sec:Model}. Section \ref{Sec:Queue} presents the Multiuser MPT queuing model. Section \ref{Sec:Results} presents the results, including the validation of the queueing model. Finally, the main conclusions of this  work are summarized and the future research lines are stated.

%
%

\section{System Model and Assumptions} \label{Sec:Model}

A network consisting of a multiple-antenna AP and $N\in[1,\infty)$ single-antenna nodes located a single hop away from the AP is considered. The AP is equipped with $M$ antennas, allowing it to create up to $\min(N,M)$ simultaneous beams and transmit a different packet in each by using a multiuser beamformer. The packets included in each transmission are selected based on a per-node FIFO scheduler as detailed in Section \ref{Sec:PerNode}. Multiple transmission rates are available, but only one is used at each transmission, and is picked based on the Channel State Information (CSI) provided by the selected nodes. Despite this rate selection, we assume that packets can still suffer errors due to both transmitter and receiver hardware characteristics, such as clock drifts  \cite{han2009all}. Erroneous packets are retransmitted until they are successfully received.

\begin{figure}[t!!!!!!!!]
\centering
\epsfig{file=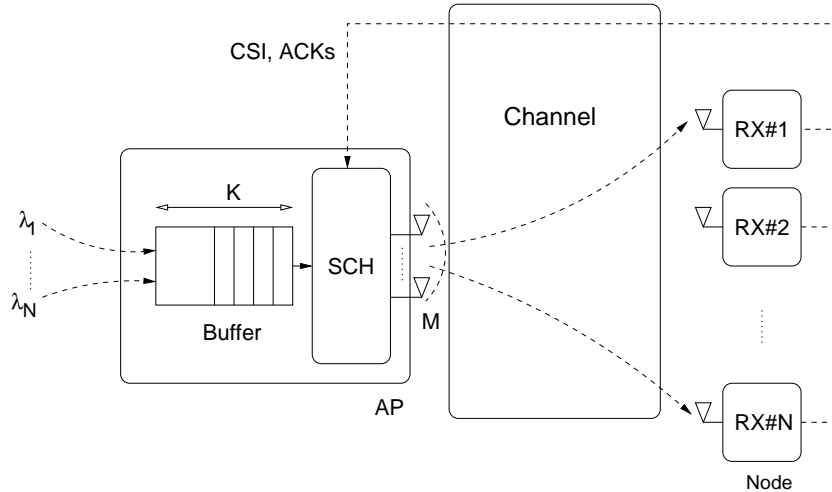,scale=0.52,angle=0}
\caption{Schematic model of an AP equipped with $M$ antennas. SCH stands for scheduler. $\lambda_n$ is the packet arrival rate directed to node $n$.}
\label{Fig:AP_model}
\end{figure}

Packets of a constant length of $L_d$ bits destined to the $N$ single-antenna nodes arrive to the AP according to an aggregate Poisson arrival process of rate $\lambda$, containing independent and identically distributed shares of traffic per node.

Given that we have to keep track of the order in which packets arrive to the AP to apply the per-node FIFO packet scheduling, a single virtual finite-buffer of size $K$ is considered, and therefore, the finite-buffer space is fully shared by all arriving packets. Moreover, as traffic differentiation between nodes is not considered, compared to the use of $N$ different queues of size $\lceil K/N \rceil$, a single shared buffer of size $K$ is optimal in terms of minimizing the packet losses due to buffer overflow \cite{kamoun1980analysis}. A detailed model of the AP architecture is shown in Figure \ref{Fig:AP_model}, including the single virtual shared buffer.

Finally, it should be noted that the considered system does not take advantage of the multiuser diversity as, at each transmission, the AP only requests the CSI of the selected users. In order to consider multiuser diversity, the AP would have to request the CSI for all the nodes with packets waiting for transmission at the AP, based on which it could select the most appropriate nodes for transmission. However, if the number of nodes is large, the time required to obtain the CSI from all nodes may be very large, specially given that the CSI information is requested and transmitted at the lowest rate. In this case, the time needed to collect and process the CSI for all nodes can be longer than the time required for both collecting the CSI and transmitting the corresponding packets for a few selected nodes, even if that is done at the lowest transmission rate. A second consideration is that with sporadic traffic, the time between two transmissions from the AP to the same node may be long, which makes it difficult to reuse previously stored CSI.

\subsection{Per-node FIFO Packet Scheduling}\label{Sec:PerNode}

At every transmission opportunity, the AP starts by constructing what we refer to as a \textit{space-batch}, containing up to $M$ packets directed to different destinations. The space-batch is constructed on a FIFO basis, however, once a packet destined to a certain node is placed in the space-batch, all subsequent packets directed to that node are skipped and left for future space-batches. This is because each node is only equipped with a single antenna and cannot receive more than one packet at a time, and therefore, there can be at most one packet per destination in every space-batch.

The scheduling of space-batches takes place immediately after the completion of each transmission if the queue is not empty. Otherwise, the AP will wait until a new arrival enters the queue, immediately after which a space-batch containing only one packet will be constructed.

When there are two or more packets present in the queue after a transmission, there may be multiple packets among them destined for the same node. Therefore, only a subset of the packets in the queue might be eligible for transmission in a single space-batch. Let $\xi$ be the number of eligible packets (destined to distinct destinations) in the queue immediately after a given transmission. Then the size of the next space-batch scheduled after that transmission is given by:

\begin{equation}\label{Eq:batch_policy}
    m = \max(1,\min(\xi,s_{\max})) = \left \{\begin{array}{lr}
			1 	~	& ~	\xi = 0 \\
			\xi ~ & ~	1 < \xi \leq s_{\max}\\
			s_{\max}	~	& ~	s_{\max} < \xi \\
          \end{array}\right.
\end{equation}
where $s_{\max}\leq M$ is a system parameter indicating the maximum number of spatial streams allowed to be sent at each transmission. Note when $\xi=0$, the next space-batch will be scheduled as soon as a packet becomes available and will always contain exactly one packet.

One of the key properties of this scheduler is that it does not discriminate between different destinations based on their channel status. Therefore, given that the buffer space is shared by all nodes, all arriving packets observe the same blocking probability, which means that destinations with a higher traffic load will have a higher chance of having packets presents in the queue and consequently, will be able to transmit more frequently. However, not choosing the destinations based on the their channel status may be detrimental in terms of the overall system performance as the system will not always transmit at the highest possible rate. To mitigate this problem, the $s_{\max}$ parameter can be used to reduce the number of parallel transmissions, as it will increase the SNR of each transmitted spatial stream and, furthermore, it will improve the performance of the multiuser beamforming, thus increasing the chances to transmit at larger rates without altering the fairness of the scheduler.

\subsection{Frame Structure}

Let $\mathcal{S}$ be the set of nodes for which a packet is included in the current space-batch. After constructing the space-batch, the AP broadcasts the identity of the nodes in $\mathcal{S}$. This is immediately followed by a series of training sequences from the AP to the selected destinations. Based on the training sequences, each node in $\mathcal{S}$ reports its estimated CSI back to the AP. The AP will use the received CSI to form the required number of beams and to choose the appropriate transmission rate, as will be detailed shortly.

The general structure of a transmitted frame is depicted in Figure \ref{Fig:Space_batch}. It consists of five parts, transmitted in the order presented below:

\begin{itemize}
 \item \emph{Preamble and space-batch information (of length $L_{\text{sb}}$ bits)}: Contains the initial preamble used to synchronize all receivers and the headers required to inform those nodes that have been selected for receiving a packet in the next space-batch.

 \item \emph{Training Sequences (of length $M \times L_{\text{tr}}$ bits)}: Required for estimating the CSI between each of the $M$ antennas at the AP and the receiving antenna at each selected node, and used to calculate the beamforming vectors.

 \item \emph{CSI feed-back (of length $m \times L_{\text{CSI}}$ bits)}: Used for nodes to report their estimated CSI to the AP.

 \item \emph{Data Packet (of length $L_{d}$ bits)}: Includes the space-batch data packets.

 \item \emph{ACKs (of length $m \times L_{\text{ACK}}$ bits)}: Used for nodes to notify the correct reception of the data packet.
\end{itemize}

\begin{figure}[tt!!!!!!!!]
\centering
\epsfig{file=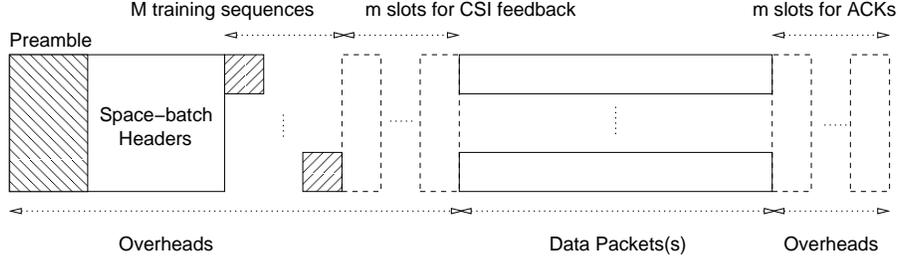,scale=0.52,angle=0}
\caption{Frame structure of a Space-batch for an MPT system where the AP has $M$ antennas. Observe how the overheads increase with the number of antennas and the number of simultaneous packets transmitted. The same frame structure as presented in \cite{bellalta2011buffer} has been considered.}
\label{Fig:Space_batch}
\end{figure}

\subsection{Channel Model and Transmission Rate Selection}\label{SEC:rate_select}

A quasi-static fading channel that changes from space-batch to space-batch transmission is considered. Let $\Gamma_{n,m}$ denote the instantaneous Signal-to-Noise ratio (SNR) observed by node $n \in \mathcal{S}$ when $m$ spatial streams are included in the space-batch, i.e., $m = |\mathcal{S}|$. The instantaneous SNR observed by a node is assumed to be independent of that of the other nodes. Considering that a ZF beamforming is used, and that the fadings are independent in both time and space at each transmission, the received SNR at each node is assumed to follow a $\chi^2$-distribution with $l = 2 \times (M - m + 1)$ degrees of freedom \cite{paulraj2003introduction}. The cumulative distribution function for {$\Gamma_{n,m}$} is therefore given by:

\begin{align}\label{Eq:ZF_SINR}
   F_{\Gamma_{n,m}}(\gamma) = \Pr\{\Gamma_{n,m} \leq \gamma\} = 1-\sum_{k=0}^{M-m}{\frac{1}{k!}\left( \frac{\gamma}{\bar{\Gamma}_{n,m}} \right)^{k} e^{-\frac{\gamma} {\bigghost \bar{\Gamma}_{n,m}}}}
\end{align}
where $\bar{\Gamma}_{n,m}$ is the average SNR observed by node $n$ when a space-batch of size $m$ is transmitted and depends only on the pathloss between the AP and node $n$. Note that $\bar{\Gamma}_{n,m}= \bar{\Gamma}_{n,1}/m$, since the transmission power is equally divided between the $m$ parallel streams.

Upon receiving the CSI feedback, which is assumed to be ideal, the AP picks a transmission rate based on the SNR values calculated for every selected node. The transmission rates are chosen from a finite set of values $\mathcal{R}=\{r_1,\ldots,r_R\}$, where $r_1 < \cdots < r_R$. The transmission rate for the $n$-th spatial stream, $\hat{r}_n$, is chosen to be $r_i$ if $\Gamma_{n,m}$ falls in the range $(\gamma_{i},\gamma_{i+1}]$, where $\{\gamma_j\}_{j=0}^{R+1}$ are predetermined thresholds. Once the transmission rate at which each spatial stream can be transmitted is known, the AP chooses the smallest rate, $\hat{r}=\min\{\hat{r}_n\}_{n\in \mathcal{S}}$, and uses it as the transmission rate for the whole space-batch. It is assumed that the probability of a space-batch suffering channel errors is negligible if the proper $\hat{r}$ is used. However, as it will be detailed in the next subsection, errors due to hardware characteristics can still happen.

It should be noted here that the control parts of the frame are all transmitted omni-directionally at $r_1$. Then, the duration of the control part of the frame is independent of the chosen transmission rate, $\hat{r}$, and is given by:

\begin{equation}\label{Eq:sb_duration}
	T_c(m)=\frac{L_{\text{sb}}+M\cdot L_{\text{tr}} + m\cdot L_{\text{CSI}} + m\cdot L_{\text{ACK}}}{r_1}
\end{equation}

The duration of a frame depends on the number of antennas and the transmission rate $\hat{r}$ chosen for the space-batch transmission, hence:

\begin{equation}\label{Eq:sb_duration}
	T(m,\hat{r})= T_c(m) + \frac{L_{d}}{\hat{r}}
\end{equation}

\subsubsection{Packet Errors}

Apart from channel conditions, packet errors can be caused by other factors such as hardware characteristics \cite{han2009all}. Since these errors do not depend on the channel conditions, we assume them to homogeneously affect all users.

Let $p_e$ be the probability that a packet suffers transmission errors. Then, the probability that $y$ of the $m$ packets included in a space-batch contain errors is given by

\begin{align}\label{Eq:psi_ym}
    \psi_{y|m}=\binom{m}{y}{p_e^{y}(1-p_e)^{m-y}}
\end{align}

\subsection{An Example of the System Operation}

\begin{figure*}[tt!!!!!!!!]
\centering
\epsfig{file=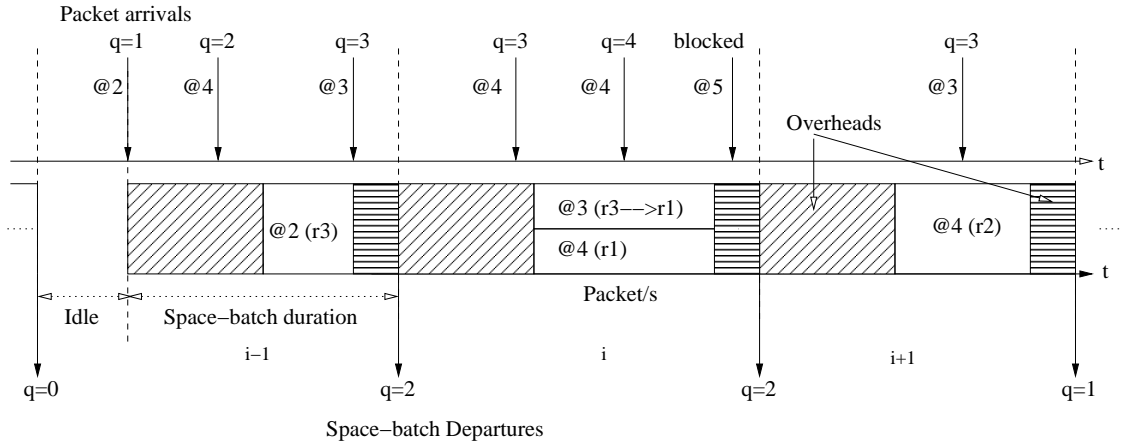,scale=0.60,angle=0}
\caption{Temporal evolution of the AP's queue with $M=2$ antennas. The $q$ values are the queue state (number of frames in the queue) just after a frame arrival or a
space-batch departure. The notation $@x$ refers to the packet destination. Observe that the duration of each transmission is proportional to the transmission rate considered. This example extends the one presented in \cite{bellalta2011buffer} by considering multiple transmission rates.}
\label{Fig:Temporal}
\end{figure*}

In Figure \ref{Fig:Temporal}, a specific example of the system operation is shown for $M=2$ antennas and a buffer of capacity $K=4$ packets, in a system with transmission rates $\mathcal{R}=\{r_1,r_2,r_3\}$. The $(i-1)$-st space-batch comprises a single packet as the transmission is scheduled as soon as a new packet arrives to the AP and it can be transmitted at rate $r_3$. During the $(i-1)$-st space-batch transmission, two packets, one directed to node $4$ and the other to node $3$, arrive to the AP and are buffered and assembled together in the $i$-th space-batch after the $(i-1)$-st space-batch transmission is completed. Based on the CSI received from nodes $3$ and $4$, rates $\hat{r}_3 = r_3$ and $\hat{r}_4 = r_1$ can be supported. The space-batch is then transmitted at rate $\hat{r} = r_1 = \min \{r_3,r_1\}$. Similarly, during the $i$-th space-batch transmission two more packets directed to node $4$ arrive to the queue, as well as one directed to node $5$, which is blocked because there is no free space in the buffer. Observe that, when the $(i+1)$-st space-batch is scheduled, there are only two packets in the transmission buffer and both are directed to node $4$. In this situation, only one packet can be transmitted, which in our example is done at rate $r_2$.

%
%

\section{Analytical Model} \label{Sec:Queue}

In this section we present the analytical model. In Table \ref{Tbl:nomenclature}, we introduce the main notation used to build the model.

\begin{table}[t!!!!]
    \centering
    \begin{small}
    \begin{tabular}{c|l}
        Variable & Description \\
        \hline
        $\lambda$ & Aggregate packet arrival rate \\
        $N$ & Number of STAs\\
        $M$ & Number of antennas at the AP\\
        $K$ & AP buffer size\\
        $R$ & Number of transmission rates \\
        $p_e$ & Packet transmission error probability \\
        $s_{\max}$ & Maximum number of packets that can be sent at each transmission  \\
        $m$ & Current space-batch size \\
        $y$ & Number of erroneous packets in a space-batch transmission \\
        $T(m,\hat{r})$ & Duration of a space-batch comprising $m$ packets and transmitted at rate $\hat{r}$\\
        $V[T(m,\hat{r})]$ & Number of packet arrivals in $T(m,\hat{r})$\\
        $s(i)$ & Maximum space-batch size when there are $i$ packets in the queue \\
        $p_{m|i}$ & Probability to transmit a space-batch of $m$ packets when there are $i$ packets in the queue\\
        $\psi_{y|m}$ & Probability that $y$ of the $m$ transmitted packets contain errors \\
        $\phi_{r_l|m}$ & Probability that a space-batch comprising $m$ packets is transmitted at rate $r_l$\\
        \hline
    \end{tabular}
    \end{small}
\caption{Notation used in the analytical model.}\label{Tbl:nomenclature}
\end{table}

\subsection{Queueing Model}  \label{Sec:QueueModel}

Poisson arrivals of rate $\lambda$ and a general service time distribution are considered. The buffer has a size of  $K$ packets and no extra space is considered for the packets in service, i.e., the packets included in a space-batch transmission remain stored in the queue until their transmission is completed.

As explained before, the scheduling of a space-batch takes place immediately after each departure, if the queue is not empty. Here a {\it departure} instant refers both to the moments at which all packets from a successfully transmitted space-batch are purged from the queue and to the end of channel outage periods. If just after a departure the queue is empty, the AP will wait until there is a new arrival, at which instant a new transmission containing a single packet is scheduled.

Let $q_k$ denote the queue occupancy at the end of the $k$-th interdeparture epoch. If $m_k$ is the  number of transmitted packets, $y_k$ the number of packets suffering errors, and $\hat{r}(k)$ the transmission rate at which the space-batch is transmitted during the $k$-th epoch, then $q_k$ evolves according to the following recursion:
\begin{equation}\label{Eq:queue_recursion}
    q_{k} = \min\left\{V[T(m_{k},\hat{r}(k))]+ q_{k-1}-m_{k}+y_{k},K-m_{k}+y_{k}\right\}
\end{equation}
where $V[T(m_{k},\hat{r}(k))]$ is the number of packet arrivals during the $k$-th inter-departure epoch. Therefore, $V[T(m_{k},\hat{r}(k))]+ q_{k-1}-m_{k}+y_{k}$ is the number of packets that would be present in the queue at the end of the $k$-th epoch if the buffer had infinite capacity, and $K-m_{k}+y_{k}$ is the maximum possible queue occupancy immediately after a departure for a finite buffer. Notice that for very small values of $K$, the queue occupancy after a departure can be lower than $s_{\max}$, which limits the size of the next space-batch by design to a lower value than its maximum. To avoid this situation, the queue size has to be at least $K\leq 2 s_{\max}$.

In order to find analytical formulation for key performance metrics such as
delay and throughput, the steady-state queue occupancy probabilities need to be calculated. To derive these probabilities, first, the steady-state distribution for the queue states immediately after departures, $\piD$, is derived using a discrete-time embedded Markov chain. Then the PASTA (Poisson Arrivals See Time Averages) property of the Poisson arrivals is applied to find the occupancy distribution at \textit{arbitrary times}, $\piS$, as a function of $\piD$.

In what follows, in order to make a clear distinction between the aforementioned two different steady-state probabilities, we define two different sets of states and a corresponding terminology for their probability distribution as follows:

\begin{itemize}
  \item {\bf queue state at departure instants:} number of packets stored in the queue immediately after a departure. Hereafter, the steady-state probability distribution for these states is referred to as the {\it departure distribution}. This is what was denoted above by the row vector $\piD$.
  \item {\bf queue state at arbitrary times:} number of packets stored in the queue at any arbitrary time. The steady-state probability distribution for these states will simply be referred to as the {\it steady-state distribution} of the queue. This is what was denoted above by the row vector $\piS$.
\end{itemize}

\subsubsection{Blind Estimation of the Space-Batch Size Distribution} \label{Sec:prob_m}

Let $p_{m|q}$ be the probability that $m$ packets are scheduled given that $q$ packets were stored in the queue at the last departure instant prior to the space-batch construction. This probability is given by
\begin{equation}\label{Eq:p_event_model}
p_{m|q} =
    \left\{
        \begin{array}{lr}
            1,&~~~m=1, q=0 \\
            \Pr\{\Xi_q =m\},&~~~ m<s_{\max}, q \geq m\\
            \sum_{\xi = s_{\max}}^N \Pr\{\Xi_q =\xi\},&~~~ m = s_{\max}, q \geq m\\
            0, & \text{otherwise}
        \end{array}
    \right.
\end{equation}
where $\Xi_q$ is the random variable denoting the number of packets eligible for transmission in a single space-batch, among the $q$ packets present in the queue. Notice that in the case that the queue is empty, i.e., $q=0$, the next space-batch will always contain only a single packet, i.e., the first packet arriving to the queue, as can be seen in the first line of (\ref{Eq:p_event_model}).

Each of the $q$ packets in the queue can be destined to any of the $N$ different nodes in the network, and therefore there are a total of $N^q$ different possible queue arrangements in terms of destination representation. To compute the probability $\Pr\{\Xi_q = \xi\}$, we need to find the fraction of such arrangements for which there are exactly $\xi$ different destinations represented. Consider one such favorable arrangement with exactly $\xi$ nodes represented. Let $\bar{\bm{\mu}}_\xi=(\mu_1, \ldots, \mu_\xi)$ be a vector containing the number of packets of each of the $\xi$ destinations represented in the queue. Then $\mu_i$ need to be positive integers and $\sum_{i=1}^\xi \mu_i = q$. We define $\Psi_{q,\xi}$ to be the set containing all such vectors, i.e., $\Psi_{q,\xi} = \{\bar{\bm{\mu}}_\xi \in  \mathbb{Z}_{+}^\xi \, | \, \sum_{i=1}^\xi \mu_i = q \}$. Then corresponding to every vector in $\Psi_{q,\xi}$, we have $\text{PR}^q_{\bar{\bm{\mu}}_\xi}$ possible different queue arrangements, where
\begin{equation}
\text{PR}^q_{\bar{\bm{\mu}}_\xi}= \frac{q!}{\prod_{i = 1}^{\xi} \mu_i !}
\end{equation}
is the number of permutations of $q$ elements, partitioned into sets of $\mu_{1},\ldots, \mu_{\xi}$ repeated elements. The probability of having $\xi$ eligible packets in the queue is then given by:
\begin{equation}\label{comb_ques_sense_rate}
\pr{\Xi_q = \xi}=\dfrac{\binom{N}{\xi}
\sum_{\bar{\bm{\mu}}_\xi \in \Psi_{q,\xi}}
\text{PR}^q_{\bar{\bm{\mu}}_\xi}}{N^q}
\end{equation}
where $\sum_{\bar{\bm{\mu}}_\xi \in \Psi_{q,\xi}}
\text{PR}^q_{\bar{\bm{\mu}}_\xi}$ is the total number of possible queue arrangements for a fixed set of $\xi$ represented nodes and $\binom{N}{\xi}$ is the number of such sets.

Notice that (\ref{comb_ques_sense_rate}) can be rewritten as follows

\begin{align}
    Pr\{\Xi_q = \xi\}=\binom{N}{\xi} \xi! \stirlingtwo{q}{1,\xi}
\end{align}
where $ \stirlingtwo{a}{b,c}$ denotes the generalized Stirling Numbers of the second kind \cite{cernuschi2001combinatorial}.

It should be noted here that in our calculations above, it is assumed that, given a randomly chosen packet from the queue, the probability that it is destined to any given target node is $1/N$, which is the probability that an arriving packet is directed to that target node. However, this is not exactly true because the space-batches are constructed containing no repeated packets and every space-batch departure will reduce the diversity of the queue. This assumption greatly simplifies the analysis, however, as we will see, it does not bear any significant impact on the analytical results, which actually match the simulations quite well.

Equation (\ref{comb_ques_sense_rate}) generalizes the results in \cite{mirkovic2008theoretical} by providing a single expression for any value of $M$, $N$ and $q$. In \cite{mirkovic2008theoretical} only a single case is provided for $q=4$, $N=4$ and $M=4$. In this case, since $M = N$, $p_{m|q}=\pr{\Xi_q = m}$ for $q>0$, which for $q = 4$ and different values of $m$ is given by:

\begin{eqnarray}
p_{1|4}&=& \frac{4}{4^{4}} \left(\frac{4!}{4!}\right)\\
p_{2|4}&=& \frac{\binom{4}{2}}{4^4}\left({\frac{4!}{1!3!}+\frac{4!}{2! 2!}+\frac{4!}{3! 1!}}\right)\\
p_{3|4}&=& \frac{\binom{4}{3}}{4^4}\left({\frac{4!}{1! 1! 2!}+\frac{4!}{1! 2! 1!}+\frac{4!}{2! 1! 1!}}\right)\\
p_{4|4}&=& \frac{1}{4^{4}}\left(\frac{4!}{1! 1! 1! 1!}\right)
\end{eqnarray}

\subsubsection{Distribution of the Selected Rate, $\hat{r}$}

Based on the rate selection mechanism explained in Section \ref{SEC:rate_select} and the SNR distribution indicated by (\ref{Eq:ZF_SINR}), the probability that a given node $n$ has a feasible rate $r_i$ is given by
\begin{equation}
\theta_{n,i|m}=\Pr\{\hat{r}_n = r_i~|~m\} = F_{\Gamma_{n,m}}(\gamma_{i+1})-F_{\Gamma_{n,m}}(\gamma_i)
\end{equation}
where $m$ is the number of {packets} included in the space-batch. Given the set of destinations, $\mathcal{S}$, included in the next space-batch transmission, a rate $\hat{r} = r_i$ will be selected if, based on the CSI feedback from those nodes in $\mathcal{S}$, the smallest feasible rate is $r_i$, i.e., $\min\{\hat{r}_n\}_{n\in\mathcal{S}} = r_i$. Therefore,  $\phi_{r_i}(\mathcal{S})$, the probability that for a given space-batch composition, $\mathcal{S}$, the rate $r_i$ is chosen, is given by:
\begin{small}
\begin{eqnarray}
	 \phi_{r_i}(\mathcal{S}) &=& \Pr \left\{ {\min\{\hat{r}_n\}_{n\in\mathcal{S}} = r_i} \right\}\nonumber\\
	 &=&\prod_{n \in \mathcal{S}}\Pr \{ \hat{r}_n \geq r_i~|~m\}-\prod_{n \in \mathcal{S}}\Pr\{\hat{r}_n \geq r_{i+1}~|~m\}\nonumber\\
	 &=&\prod_{n \in \mathcal{S}}{\left(\sum_{j=i}^{R}{\theta_{n,j|m}}\right)}-\prod_{n\in\mathcal{S}}{\left(\sum_{j=i+1}^{R}{\theta_{n,j|m}}\right)}
\end{eqnarray}
\end{small}
where the second equality is due to the assumed independence of the observed SNR values for different nodes and can be interpreted as the probability that all the nodes in $\mathcal{S}$ have rates no smaller than $r_i$ but not all of those rates are strictly larger than $r_i$.

\subsubsection{Distribution at departure epochs, $\piD$}

The departure probability distribution, $\piD$, is computed by solving the linear system:
\begin{equation}\label{Eq:EMC}
  \piD = \piD \Pmat
\end{equation}
together with the normalization condition:
\begin{equation}
\piD \mathbf{1}^{T} = 1
\end{equation}
where $\Pmat$ is the probability transition matrix of the
embedded discrete-time Markov chain of the occupancy of the batch-service queue, immediately after departure instants, with each element $p_{i,j}$, $i,j\in[0,K]$, representing the probability to move from state $i$ to state $j$.

In this chain, transitions occur at departure instants, i.e., immediately after the complete transmission of a frame or at the end of a channel outage period, and the states represent the queue occupancy immediately after a departure. The $p_{i,j}$ transition probabilities can be viewed as
\begin{equation}
p_{i,j} = \pr{q_k = j|q_{k-1} = i}
\end{equation}
for any $k$, where $q_k$ and $q_{k-1}$ are related according to the recursion in (\ref{Eq:queue_recursion}). As it can be seen in the recursion, $p_{i,j}$ not only depends on the size of the transmitted space-batch, but also on the time it takes to transmit the corresponding frame, which depends on the chosen rate $\hat{r}$, which in turn depends on $\mathcal{S}$, the composition of the space-batch.

Let $p_{i,j}(m,y,r)$ denote the conditional transition probability given that at state $i$ a space-batch of size $m$ packets is transmitted at rate $r$ and there are $y$ erroneous packets. This probability depends on the value of $V[T(m,r)]$, the random variable representing the number of arrivals during the transmission of the $m$ packets sent at rate $r$. For any state $i$ in the chain, the last reachable state is $j=K-m+y$. Therefore,
\begin{equation}\label{Eq:transition_pr_conditional}
    p_{i,j}(m,y,r)= \left\{
        \begin{array}{lr}
            \pr{V[T(m,r)] = j - (i-m+y)}, & ~  j < K-m+y\\
            \pr{V[T(m,r)] \geq j - (i-m+y)}, & ~ j = K-m+y
        \end{array}
        \right.
\end{equation}
where $i\in \left[0,K\right]$, $j \geq i-m+y$, and $m\leq s(i)$, with the function $s(i)=\max(1,\min(i,s_{\max}))$ defined as the maximum possible size of next space-batch when at the end of last departure there are $i$ packets in the queue. For all other values of $i,j$, we have $p_{i,j}(m,y,r) = 0$. Note that departing at state $j = K-m+y$ means that the queue has been containing $q = K$ packets just before the departure, and therefore, some arrivals have possibly been blocked. For all other reachable states from state $i$, the queue has had room for more packets just before the departure and therefore no arrivals could have been blocked.

For Poisson arrivals of rate $\lambda$, the number of arrivals during $T(m,r)$ has in general the following distribution:
\begin{equation}\label{Eq:arrival_per_cycle_pdf}
	    \pr{V[T(m,r)] = v}=\int_{0}^{\infty}{e^{-\lambda t}\frac{(\lambda t)^v}{v!} f_{T(m,r)}(t)}dt
\end{equation}
where $f_{T(m,r)}(t)$ is the probability density function of $T(m,r)$. In our case, since given $m$ and $r$ the frame duration $T(m,r)$ is constant, it can be simplified to:
\begin{equation}\label{Eq:arrival_per_cycle_pdf_det}
		\pr{V[T(m,r)] = v} = e^{-\lambda T(m,r)}\frac{(\lambda T(m,r))^v}{v!}
\end{equation}

For any feasible state pair $(i,j)$, i.e., $i\in \left[0,K\right]$ and $j \in \left[i-m+y,K-m+y\right]$, from (\ref{Eq:transition_pr_conditional}) and (\ref{Eq:arrival_per_cycle_pdf_det}), we have:
\begin{equation}\label{Eq:p_ij}
    p_{i,j}(m,y,r)=\left \{\begin{array}{lr}
			\displaystyle{e^{-\lambda T(m,r)}\frac{(\lambda T(m,r))^v}{v!}} ,	 & ~ j < K-m+y \\
			\displaystyle{1-\sum_{z=i-m+y}^{K-m+y-1}{p_{i,j}(m,y,r)}} ,
& ~ j = K-m+y
          \end{array}\right.
\end{equation}

To calculate the unconditional transition probabilities, $p_{i,j}$, we need to average $p_{i,j}(m,y,r)$ over all possible values of $m$, $y$ and $r$, i.e.,
\begin{equation}\label{Eq:transition_probs}
     p_{i,j}=\sum_{m=1}^{s(i)}{p_{m|i}\sum_{y=0}^{m}{\psi_{y|m}\sum_{l=1}^{R}{\phi_{r_l|m}p_{i,j}(m,y,r_l)}}}
\end{equation}
where $p_{m|i}$ is given by (\ref{Eq:p_event_model}), with $q=i$, and $\psi_{y|m}$ by (\ref{Eq:psi_ym}). $\phi_{r_l|m}$ is the probability that the rate $r_l$ is chosen for a space-batch, given that it contains $m$ packets. This probability is given by:
\begin{equation}
    \phi_{r_l|m} =\frac{\sum_{\forall \mathcal{S},~|\mathcal{S}|=m}\phi_{r_l}(\mathcal{S})}{\binom{N}{m}}
\end{equation}
where the sum is taken over all sets $\mathcal{S}$ of cardinality $m$, i.e., containing $m$ distinct destination nodes.

\subsubsection{Distribution at arbitrary times, $\piS$}

Using the PASTA property \cite{gross1998fundamentals} of Poisson arrivals, the probability that at an arbitrary time in the steady-state the queue contains $q = k$ packets is equal to the probability that a random arrival observes $k$ packets in the queue. In other words,
\begin{equation} \label{Eq:pis1}
  \pi^s_k=\pr{q_a{(t)}=k}
\end{equation}
where $\pi^s_k$ is the $k$-th element of $\piS$, and $q_a{(t)}$ is the state of the queue observed by an arrival at time $t$. The right hand side of (\ref{Eq:pis1}) can be expanded by conditioning on $q_d(t)$, the state of the queue at the most recent departure before $t$, i.e.,

\begin{equation} \label{Eq:pis2}
  \pi^s_k = \sum_{i=0}^{k}{\pr{q_a{(t)}=k|q_d{(t)}=i}\pr{q_d{(t)}=i}}
\end{equation}

In order to calculate $\pr{q_d{(t)}=i}$, we observe that this probability can be viewed as the probability that the arrival at time $t$ happens to occur during departure state $i$, i.e., when the node is in state $i$ of the embedded Markov chain discussed in the previous subsection. Therefore, this probability is equal to the expected fraction of time that the node spends in the departure state $i$.

Let the random variable $W(i)$ denote the time spent in departure state $i$. Of this time, $T_d(i)$ seconds will be spent in transmission mode, and only if $i = 0$, an additional $I$ seconds will be spent in idle mode before entering transmission mode. Therefore:
\begin{equation}\label{Eq:epoch_duration}
    E[W(i)] = E[I] + E[T_d(i)] = \frac{1}{\lambda}\left[ 1-i \right]^+ + E[T_d(i)]
\end{equation}
where the term $\frac{1}{\lambda}\left[ 1-i \right]^+$ is nonzero only when $i=0$ and is equal to the expected time needed for the queue occupancy to reach $1$ packet. Then, the expected length of an interdeparture epoch is

\begin{align}\label{Eq:epoch_duration}
    E[W] &= \sum_{i=0}^{K}{\pi^d_i \left(\frac{1}{\lambda}\left[ 1-i \right]^+ + E[T_d(i)]\right)} \nonumber\\
    &=\frac{1}{\lambda}\pi^d_0+\sum_{i=0}^{K}{\pi^d_i E[T_d(i)]}
\end{align}
with

\begin{equation}\label{Eq:epoch_duration}
    E[T_d(i)] =\sum_{m=1}^{s(i)}{p_{m|i}{\sum_{l=1}^{R}{\phi_{r_l|m}T(m,r_l)}}}
\end{equation}

The probability $\pr{q_d{(t)}=i}$ in (\ref{Eq:pis2}) can now be calculated as follows:
\begin{equation} \label{Eq:depart_i}
  \pr{q_d{(t)}=i} = \frac{\pi^d_i E[W(i)]}{E[W]}
\end{equation}
which can be interpreted as the fraction of total time spent in departure state $i$.

The term ${\pr{q_a{(t)}=k|q_d{(t)}=i}}$ in (\ref{Eq:pis2}) is the probability that an arrival during the departure state $i$ observes $k$ packets in the queue. This probability can be viewed as the fraction of arrivals in departure state $i$ which observe $k$ packets in the queue at the moment of their arrival. The expected total number of arrivals in state $i$ is given by $\lambda E[W(i)]$. Of these, only one may observe $k<K$ frames in the queue, provided that there are enough arrivals. Let $q_{n+1}$ be the state at which the next departure will leave the queue. If the space-batch size for this departure is $m$, and it contains $y$ erroneous packets, then the queue occupancy just before this next departure is $q_{n+1}+m-y$. In order for an arrival to have observed $k<K$ packets in the queue, we need $q_{n+1} + m-y \geq k+1$ packets. Therefore, the probability that an arrival in state $i$ observes $k$ packets in the queue, given the space-batch size $m$ and there are $y$ erroneous packets, is given by:

\begin{align}\label{Eq:prob_k_observed_given_m}
    \Pr\left\{q_{n+1} \geq k+1 -m+y | q_n = i\right\}=\sum_{y=1}^{m}{\psi_{y|m}{\sum_{l=1}^{R}{\phi_{r_l|m}\sum_{j = k+1-m+y}^{K-m+y} p_{i,j}(m,y,r_l)}}}
\end{align}

The probability that an arrival in state $i$ observes $k$ packets in the queue is then given by:

\begin{align}\label{Eq:prob_i_j_dep_}
    \Pr\left\{q_a{(t)}\right.\left.=k |q_d{(t)}=i\right\} = \frac{\sum_{m=1}^{s(i)}{p_{m|i} \pr{q_{n+1} \geq k+1-m+y | q_n = i}}}{\lambda E[W(i)]}
\end{align}

From (\ref{Eq:pis2}), (\ref{Eq:depart_i}), and (\ref{Eq:prob_i_j_dep_}), the steady state queue occupancy distribution, $\piS$,  for states $0 \leq k \leq K-1$ can be computed as shown in (\ref{Eq:pis_k}).

\begin{align}\label{Eq:pis_k}
 \pi^s_k=\frac{1}{\lambda E[W]}\sum_{i=0}^{k} \pi_i^d  \left(\sum_{m=1}^{s(i)}{p_{m|i}\left(\sum_{y=1}^{m}{\psi_{y|m}\left(\sum_{l=1}^{R}{\phi_{r_l|m}\sum_{j = k+1-m+y}^{K-m+y} p_{i,j}(m,y,r_l)} \right)}\right)}\right)
\end{align}

For $k=K$, we have

\begin{equation} \label{Eq:pis_k_last}
  \pi^s_K= 1 - \sum_{i = 0}^{K-1} \pi^s_i
\end{equation}

Note that in (\ref{Eq:pis_k}), when $k=0$, for all values of $m$ and $y$ we have
$$\sum_{j = k+1-m+y}^{K-m+y} {p_{i,j}(m,y,r)}=1,$$ and (\ref{Eq:pis_k}) simplifies to:
\begin{equation} \label{Eq:pis_k_smin}
  \pi^s_0=\frac{1}{\lambda E[W]} \pi_0^d
\end{equation}
This is because in this case, during the departure state $i=0$, exactly one arrival will observe $0$ packets in the queue with probability $1$.

\subsection{Performance Metrics}

Once the $\boldsymbol{\pi}^d$ and $\boldsymbol{\pi}^s$ distributions are obtained, several performance metrics can be derived from them:

\begin{itemize}
 \item {\bf Blocking Probability:} The probability that an arriving packet to the AP is discarded because there is no space for it in the transmission queue: $P_b=\pi^s_K$
 \item {\bf System Throughput:} Number of bits that can be successfully transmitted from the AP per second: $S=\lambda(1-P_b)L_d$
 \item {\bf Average Queue Occupancy:} The average number of packets in the queue: $E[Q]=\sum_{q=1}^{K}{q\pi^s_q}$
 \item {\bf Average Response Delay:} The average delay that a packet suffers, from its entrance to the queue until it is transmitted, computed from the average queue occupancy by applying the Little's Law \cite{gross1998fundamentals}: $E[D]=\frac{E[Q]}{\lambda(1-P_b)}$
 \item {\bf Average Space-Batch Size:} Average number of packets included in the transmitted space-batches: $E[s]=\sum_{q=0}^{K}{\pi^d_q\left(\sum_{m=1}^{s(q)}{m \cdot p_{m|q}}\right)}$
\end{itemize}

%
%

\section{Results} \label{Sec:Results}

In this section, the analytical model is validated through simulations, and some insights on how the number of antennas, number of active users, channel conditions, and traffic load impact the performance of a SDMA-based Multiuser MPT system are provided. The values of the parameters used for both the simulations and the analytical model are listed in Table \ref{Tbl:parameters}.

A simulator of the described scenario has been built, from scratch, using the C++ language and based on the COST (Component Oriented Simulation Toolkit) libraries \cite{chen2001component}. The simulator accurately reproduces the system operation described in Section \ref{Sec:Model}. Therefore, by comparing the results obtained from the simulator with the ones obtained from the analytical model, we can assess the accuracy of the analytical model and observe the impact of the different assumptions used to built it. For each point, a single simulation with a duration of $1000$ seconds has been run. This duration is sufficiently long for getting confidence intervals that are not graphically visible.

In the results, we first focus on the impact of the number of users and queue size on the system performance. We then shift our focus to the effect of channel conditions by considering different SNR and $p_e$ values. Finally, we consider the case in which the traffic is heterogeneous.

\begin{table}[h!!!]
\centering
 \begin{tabular}{c|c}
  Parameter & Value \\
  \hline
  $L_{\text{sb}}$ & $256$ bits \\
  $L_{\text{tr}}$ & $64$ bits \\
  $L_{\text{CSI}}$ & $64$ bits \\
  $L_d$ & $8000$ bits \\
  $L_{\text{ACK}}$ & $64$ bits \\
  $M$ & $8$ antennas \\
  $\left\{\gamma_j\right\}$ & $\left\{10,~15,~20,~+\infty\right\}$ dB\\
  $\mathcal{R}$ & $\left\{6,~12,~18,~24\right\}$ Mbits/second \\
  \hline
 \end{tabular}
 \caption{Parameters considered for the performance evaluation}\label{Tbl:parameters}
\end{table}

\subsection{Buffer Size and Number of Users}

\begin{figure}[h!]
\centering
\subfigure[Blocking Probability]{\epsfig{file=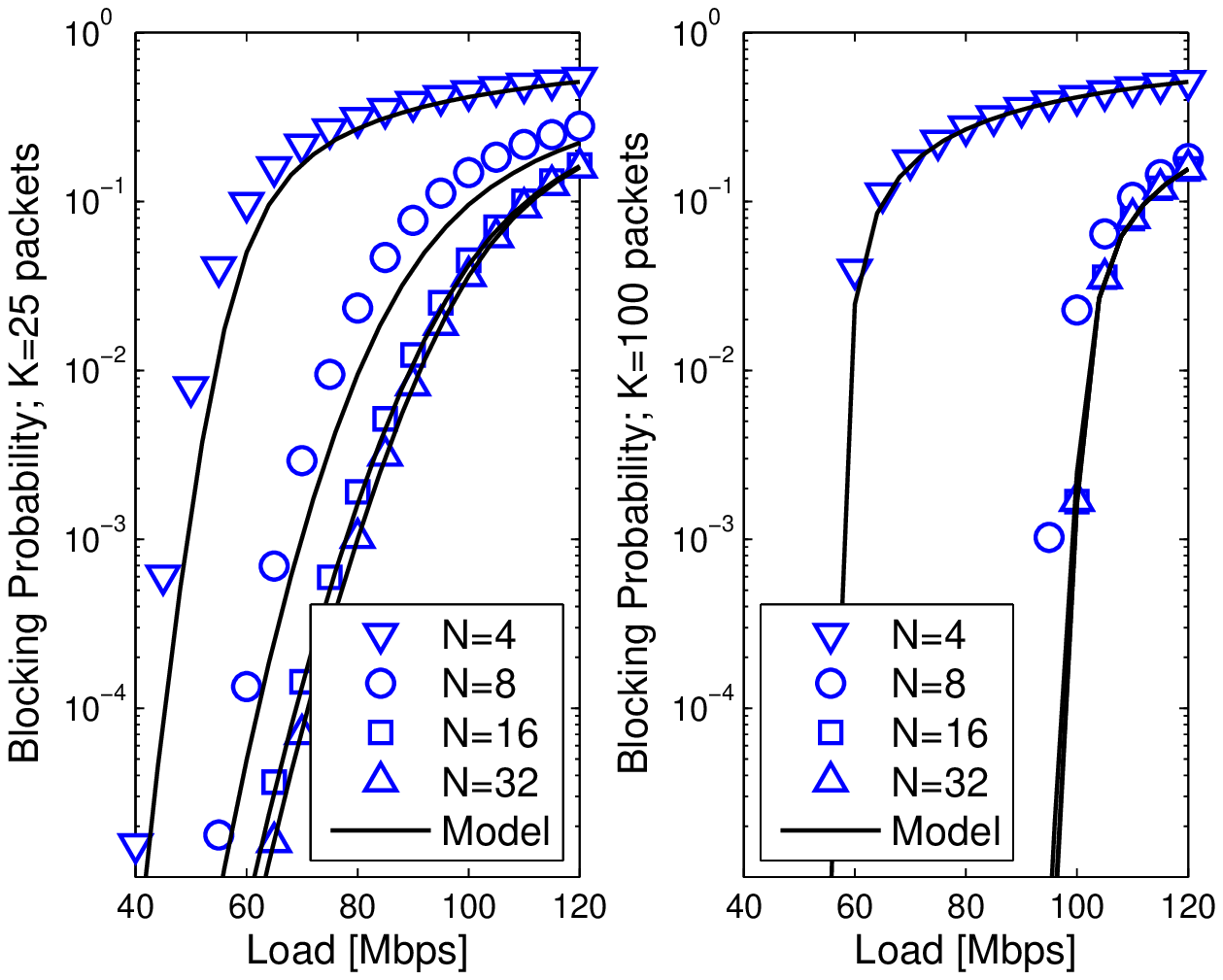,scale=0.5650,angle=0}\label{Fig:LoadKU_BP}}
\subfigure[Expected Delay]{\epsfig{file=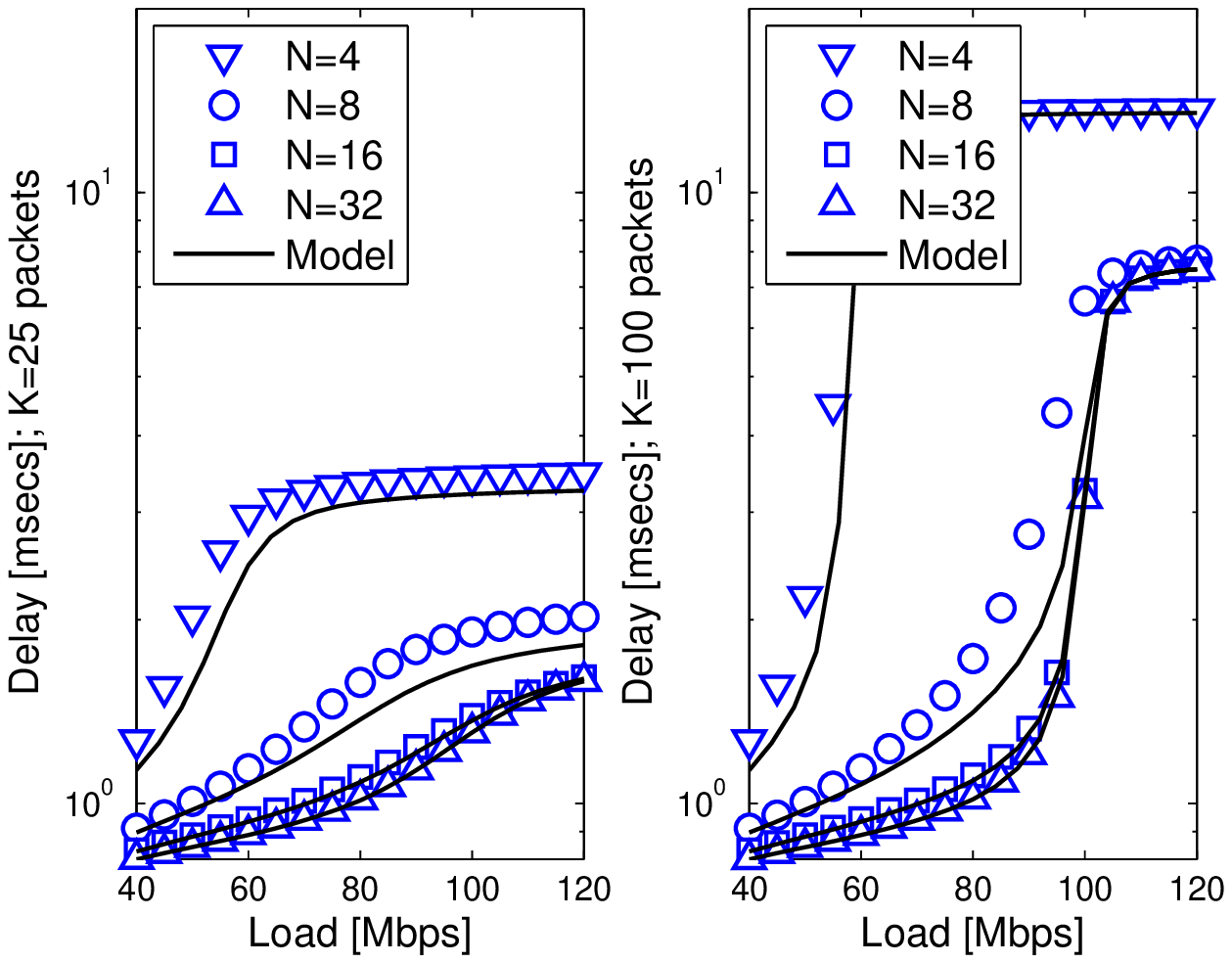,scale=0.5650,angle=0}\label{Fig:LoadKU_ED}}\\
\subfigure[Expected Space-batch size]{\epsfig{file=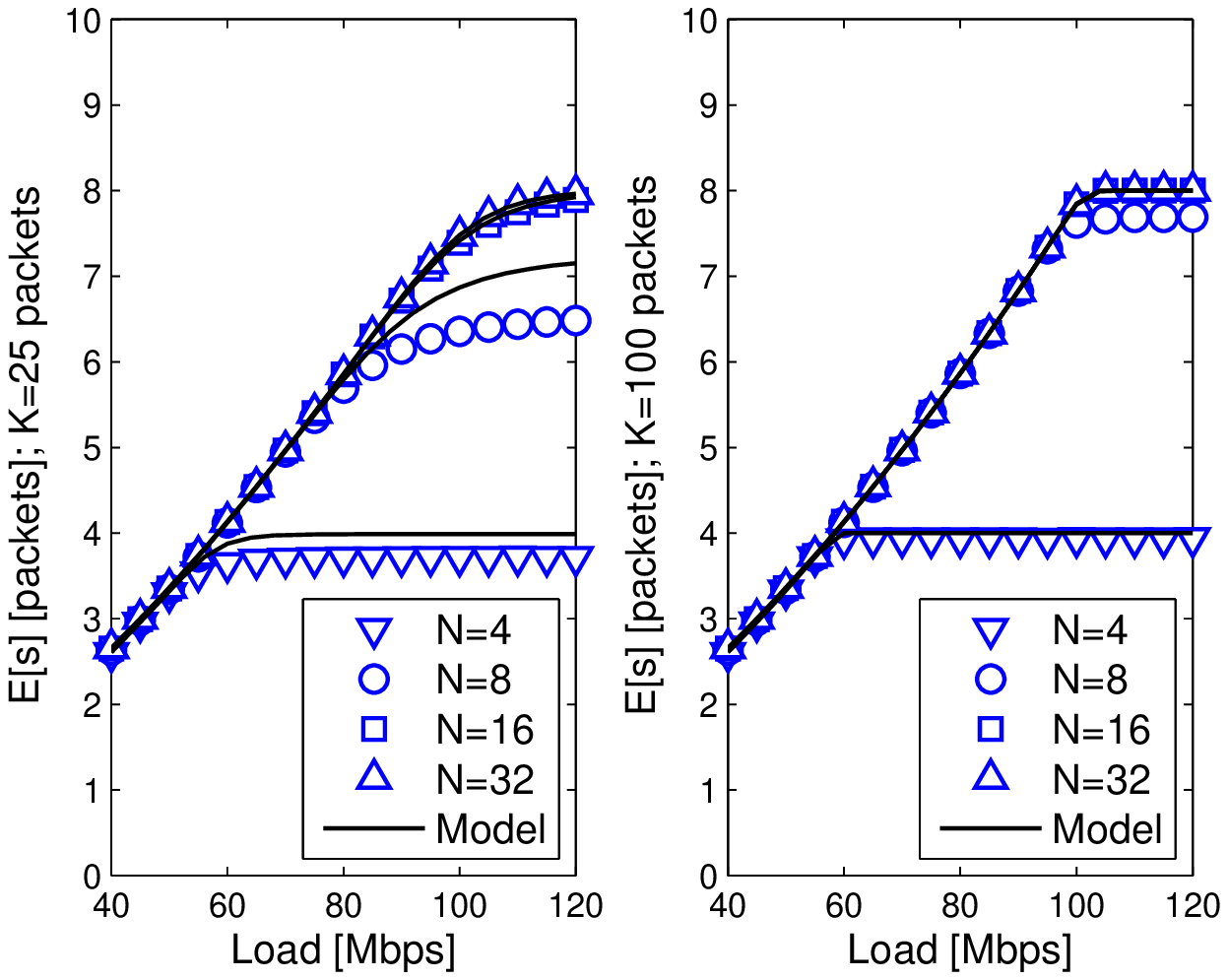,scale=0.5650,angle=0}\label{Fig:LoadKU_EB}}
\subfigure[Expected Queue Size]{\epsfig{file=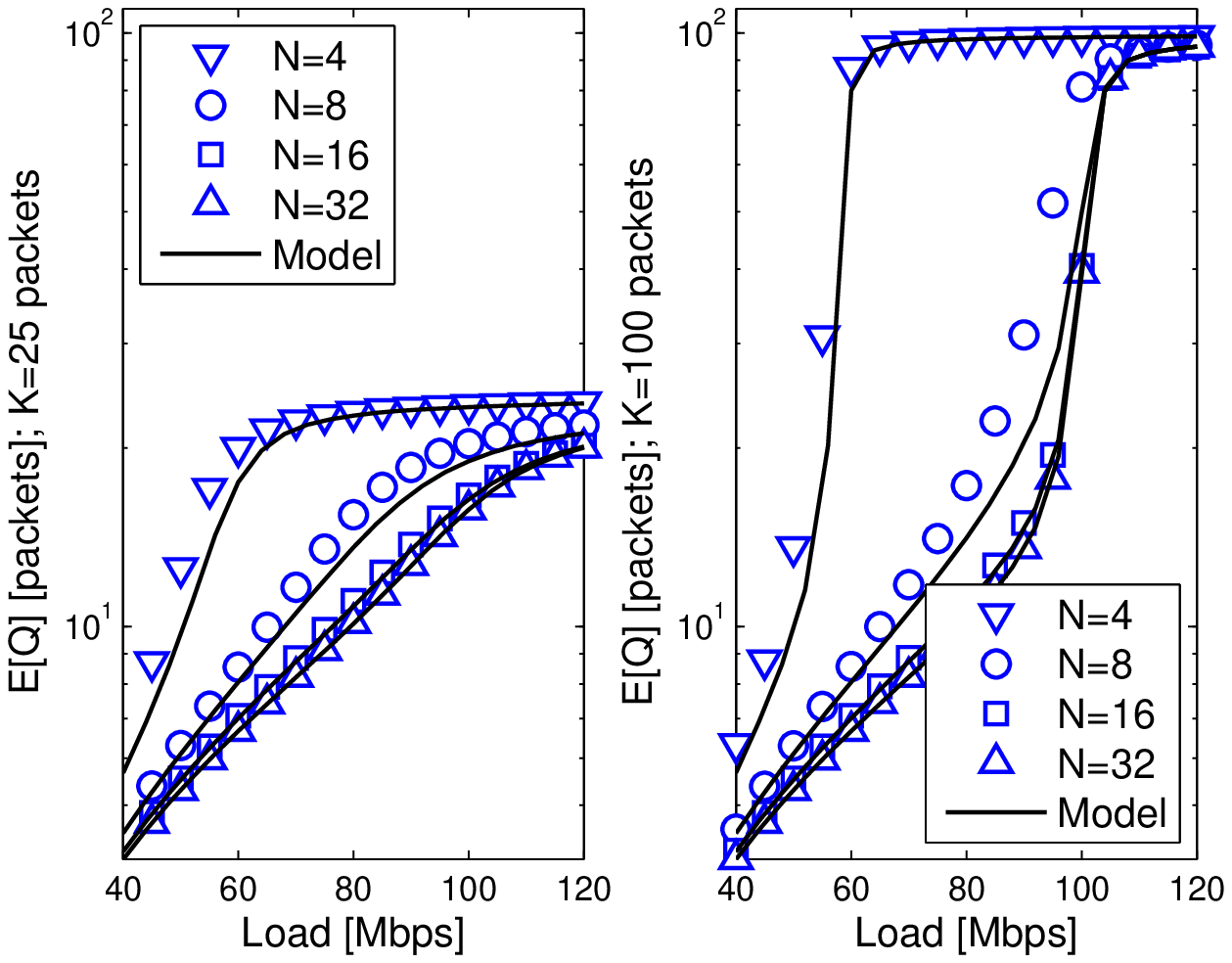,scale=0.5650,angle=0}\label{Fig:LoadKU_EQ}}
\caption{System Performance for $M=8$ antennas and different $N$ and $K$ values in ideal channel conditions.} \label{Fig:LoadKU}
\end{figure}

Figure \ref{Fig:LoadKU} shows the behavior of different performance metrics under ideal channel conditions for two values of $K$ ($K=25$ and $K=100$ packets) and four values of $N$ ($N=4$, $N=8$, $N=16$ and $N=32$ nodes) when the aggregate traffic load ($\lambda L$) increases from $40$ Mbps to $120$ Mbps. By ideal channel conditions, we refer to the case in which the AP can always transmit at the highest available transmission rate ($24$ Mbps), regardless of the number of parallel streams being transmitted, and $p_{\text{e}}=0$. Therefore, $s_{\max}$ can be set to its highest value ($s_{\max}=M=8$) to maximize the number of packets that can be included in each space-batch. We consider this scenario in order to focus on how the buffer size and the number of users affect the accuracy of the analysis. In this case, the system performance in terms of blocking probability (Figure \ref{Fig:LoadKU_BP}) and expected delay (Figure \ref{Fig:LoadKU_ED}) is only affected by the ability of the AP to schedule large space-batches (Figure \ref{Fig:LoadKU_EB}), which in turn depends on the queue occupancy (Figure \ref{Fig:LoadKU_EQ}) and the number of nodes sharing the aggregate traffic load. Obviously, in the case where there are fewer nodes than $s_{\max}$, the system performance is limited by the number of nodes.

For $K=25$, in terms of blocking probability and delay, it can be observed that as the number of nodes increases, for any given aggregate load, the system performance improves since the increased diversity in traffic makes the construction of larger space-batches more likely. For $K=100$, the same behavior can be observed, although the performance gain achieved by increasing the number of nodes is less significant. This is because a larger buffer can store more packets, and therefore, the probability that it contains packets directed to a higher number of destinations is also higher, which allows for transmission of larger space-batches even when the number of nodes is small. However, on the downside, a larger queue size results in a longer expected delay due to increased waiting time.

In terms of accuracy, the precision of the analytical model improves as the queue size grows. Since any inaccuracy between the analytical model and the simulations is due to to the blind estimation of the number of nodes represented in the queue, equation (\ref{Eq:p_event_model}) provides a better estimation of the space-batch size when the queue occupancy is higher, and there is potentially a higher number of destinations represented in the queue. The accuracy of the analytical model is also a function of the number of nodes through (\ref{Eq:p_event_model}). For $N=1$, as there cannot be any error in the estimation of the number of nodes represented in the queue, the model is exact. For $N\leq M$, the diversity that is removed from the queue at every transmission increases with $N$, and therefore, the accuracy decreases and reaches its minimum at $N=M$. Finally, for $N>M$, as $N$ increases, the analytical model becomes more accurate. This is because in this case the space-batch size is often smaller than the queue diversity, and therefore the departure has a less significant effect on the remaining diversity of the queue. This is why in Figure \ref{Fig:LoadKU}, where $M=8$, the analytical curves corresponding to $N=4$ and $N=8$ are the curves showing the highest, albeit not significant, discrepancy with their simulated counterparts.

\subsection{Heterogeneous Channel Conditions}

\begin{figure}[h!]
\centering
\epsfig{file=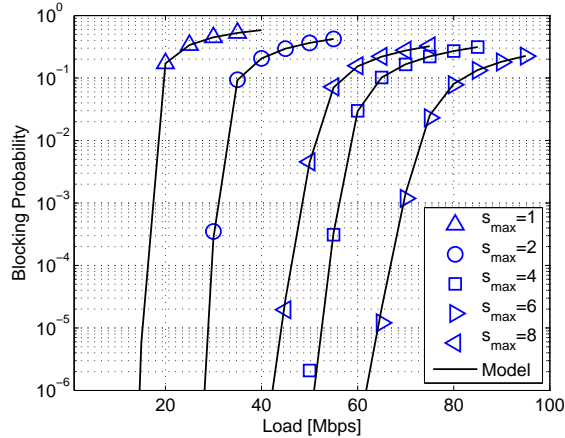,scale=0.565,angle=0}
\caption{Blocking Probability for $M=8$ and $N=16$ nodes in heterogeneous channel conditions for different $s_{\max}$ values.}
\label{Fig:HLoad_BP}
\end{figure}

In Figure \ref{Fig:HLoad_BP}, the blocking probability for $N=16$ nodes, $K=50$ packets, and different $s_{\max}$ values is plotted against different traffic loads in heterogeneous channel conditions. The $16$ nodes are distributed in the coverage area of the AP, in a way that a first group of $5$ nodes observe an average SNR equal to $25$ dBs, a second group of $5$ nodes observe an average SNR equal to $45$ dBs, and finally, a third group of $6$ nodes observe an average SNR equal to $35$ dBs. In all cases the packet error probability is $p_{\text{e}}=0$.

As can be seen in this figure, lower blocking probability can be achieved by appropriately choosing the value of $s_{\max}$, which in this specific case is $s_{\max}=6$. The optimal $s_{\max}$ value is a trade-off between the number of packets included in each space-batch and the transmission rate at which the space-batch can be transmitted. When the traffic load increases, the queue occupancy grows, and the probability to schedule larger space-batches also increases. However, as the CSI is not used for selecting neither the number nor the specific destinations to which packets are sent, the transmission rate at which space-batches are sent decreases as the number of spatial streams increases. Therefore, using the $s_{\max}$ parameter, the system can achieve better performance by trading off the maximum number of packets transmitted with the average transmission rate observed. The optimal value of $s_{\max}$ increases with the SNR observed by the nodes.

\subsection{Errors}

\begin{figure}[h!]
\centering
\epsfig{file=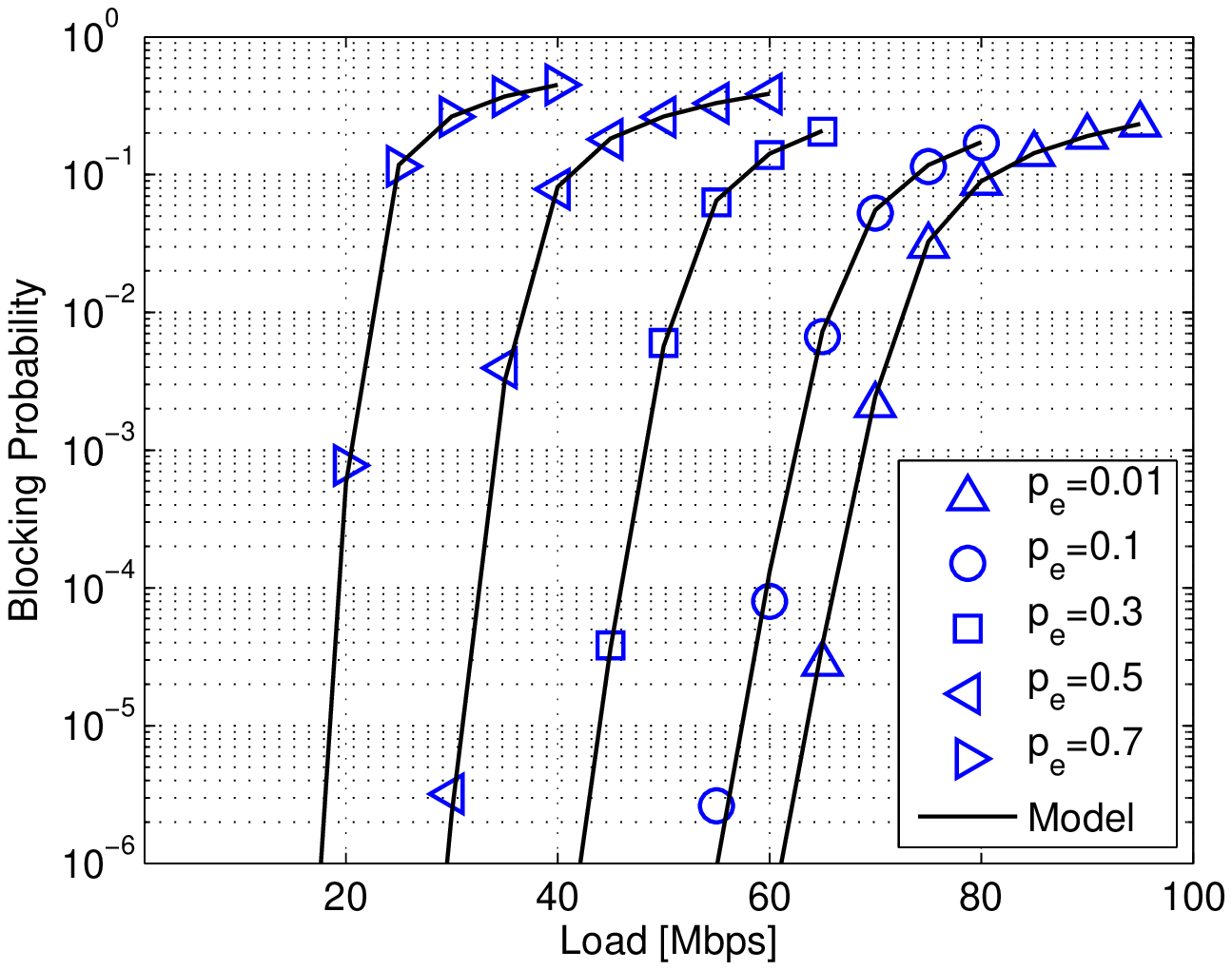,scale=0.565,angle=0}
\caption{Blocking Probability for $M=8$, $s_{\max}=6$ and $N=16$ nodes in heterogeneous channel conditions for different $p_{\text{e}}$ values.}
\label{Fig:HLoad_Errors}
\end{figure}

In Figure \ref{Fig:HLoad_Errors}, the blocking probability for $N=16$ nodes, $K=50$ packets, $s_{\text{smax}}=6$, and different $p_{\text{e}}$ values is plotted against different traffic loads in non-ideal and heterogeneous channel conditions. The $16$ nodes are distributed in the coverage area of the AP, in a way that a first group of $5$ nodes observe an average SNR equal to $25$ dBs, a second group of $5$ nodes observe an average SNR equal to $45$ dBs, and finally, a third group of $6$ nodes observe an average SNR equal to $35$ dBs.

The results show the effect of packet errors in the blocking probability. Increasing \per, the number of packets successfully transmitted in each space-bath decreases, which reduces the packet departure rate from the queue, increasing the packet blocking probability. For example, for a traffic load of $60$ Mbps, the blocking probability with $p_e=0.1$ is $10^{-4}$, which moves to $10^{-1}$ for $p_e=0.3$.

\subsection{Heterogeneous Traffic Loads}

The goal of this subsection is to show that the presented model, even though it only supports homogeneous traffic, can be used to understand the system performance when it carries heterogeneous flows with certain patterns. In detail, by concentrating the traffic in a subset of nodes, the performance in terms of blocking probability can be lower bounded by a homogeneous network of the same size, and upper bounded by another homogeneous network of the size of the subset.

The impact of having a different traffic load for each node is evaluated using the same ideal channel conditions as in the first case, considering $N=16$ users, $M=8$ antennas, $s_{\max}=8$ and $K=50$ packets. The traffic load for node $i$ is $\lambda_i=\frac{\alpha_i}{\sum_{\forall{j}}{\alpha_j}}\lambda$, with $\alpha_i$ a traffic scaling parameter that determines the fraction of traffic directed to node $i$, and $\lambda$ the aggregate packet arrival rate, which follows a Poisson process. Five different traffic profiles are considered here, as listed in Table \ref{Tbl:TP}. They are designed to evaluate how the overall system performance is affected when the fraction of traffic directed to a subset of the active nodes increases. For instance, TP2 assigns eight times more traffic to the first four nodes than to the other twelve nodes. As it will be observed, the presence of heterogeneous traffic will cause a loss on the system performance, mainly because it reduces the queue diversity and consequently, the ability to schedule large space-batches.

\begin{table}[h!!!]
\centering
 \begin{tabular}{c|c}
  TP & $\{\alpha_i,\ldots,\alpha_{16}\}$  \\
  \hline
  Hom. & $\{1,1,1,1,1,1,1,1,1,1,1,1,1,1,1,1\}$ \\
  TP1 & $\{4,4,4,4,1,1,1,1,1,1,1,1,1,1,1,1\}$ \\
  TP2 & $\{8,8,8,8,1,1,1,1,1,1,1,1,1,1,1,1\}$ \\
  TP3 & $\{16,16,16,16,1,1,1,1,1,1,1,1,1,1,1,1\}$ \\
  TP4 & $\alpha_i\sim \mathcal{U}[0,16]$, $\alpha_i \in \mathcal{R}$ \\
  \hline
 \end{tabular}	
 \caption{Traffic Profiles (TP)	}\label{Tbl:TP}
\end{table}

\begin{figure}[h!]
\centering
\subfigure[Blocking Probability]{\epsfig{file=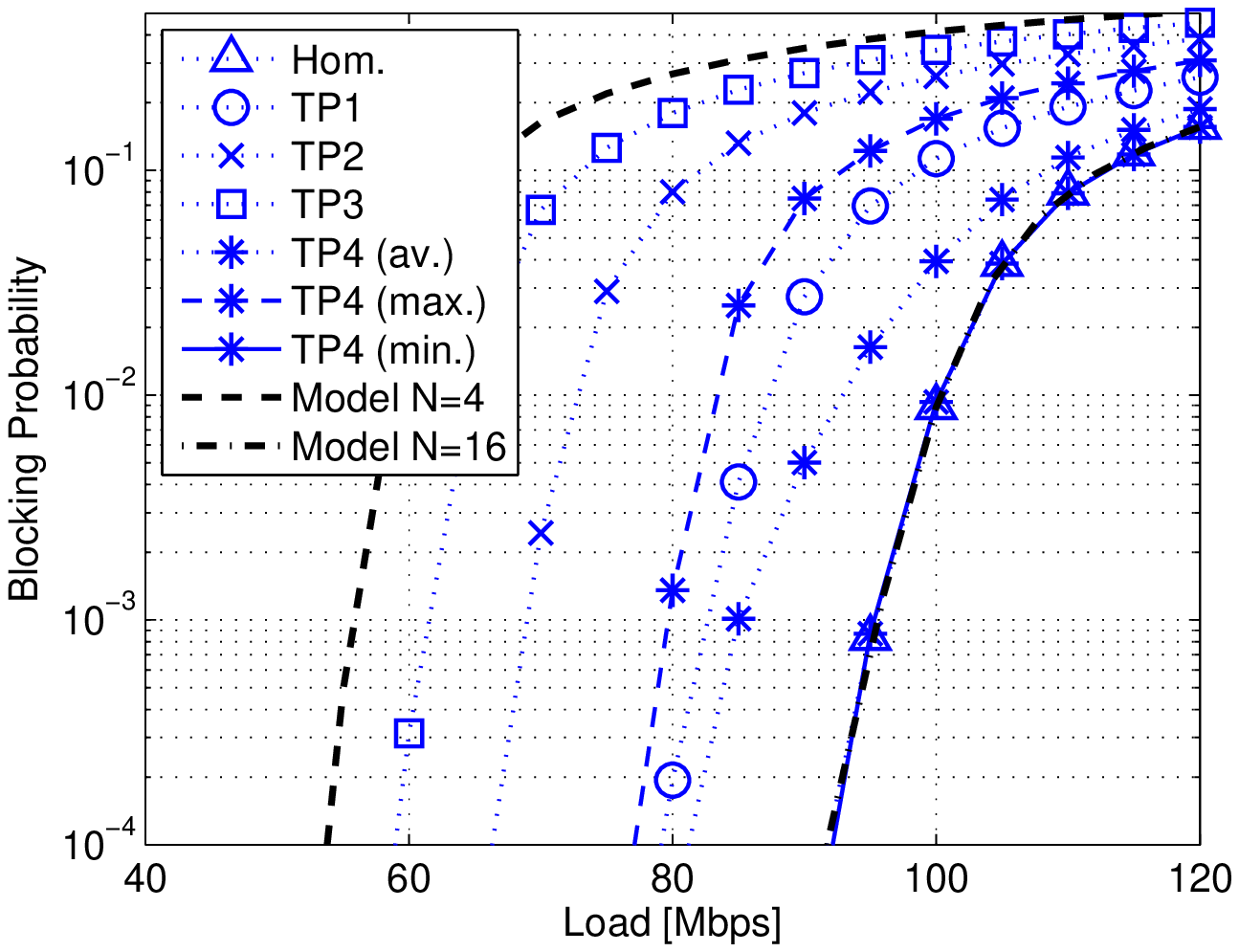,scale=0.565,angle=0}\label{Fig:HetTraffic_Pb}}
\subfigure[Expected Space-batch size]{\epsfig{file=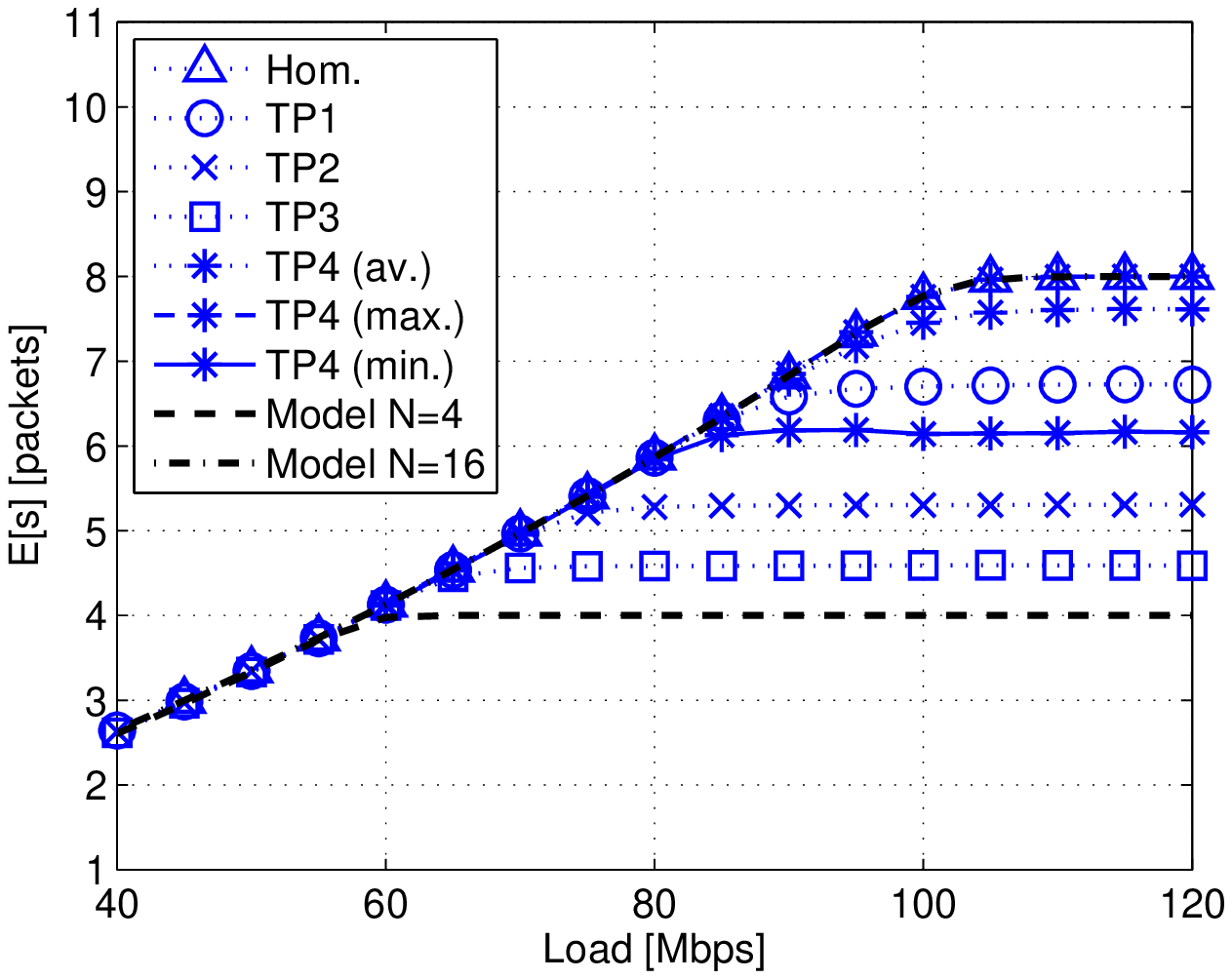,scale=0.565,angle=0}\label{Fig:HetTraffic_EB}}
\caption{Blocking Probability and Expected Space-batch size for $M=8$ antennas, $N=16$ nodes, and heterogeneous traffic in ideal channel conditions. Since TP4 is random, we have plotted $3$ curves obtained from $30$ different simulation runs. The \textit{TP4 (av.)} curve shows the average, the \textit{TP4 (max.)} curve shows the maximum and the \textit{TP4 (min.)} curve shows the minimum of the $30$ values obtained for each point.} \label{Fig:HetTraffic}
\end{figure}

Figures \ref{Fig:HetTraffic_Pb} and \ref{Fig:HetTraffic_EB} show the blocking probability and the expected space-batch size respectively for the different traffic profiles from Table \ref{Tbl:TP}, as well as the blocking probability calculated using the model for $N=16$ and $N=4$ nodes with homogeneous traffic. For all non-homogeneous traffic patterns, the plots are obtained using simulation.

It can be observed that when the aggregate traffic load is more concentrated among the four nodes indicated in Table \ref{Tbl:TP}, the blocking probability increases and tends to the performance of a network consisting of only four homogeneous nodes. This is clearly observed in Figure \ref{Fig:HetTraffic_Pb} for the case where four users have $16$ times the traffic load of each one of the remaining twelve users (i.e., TP3). This loss on performance is caused by the reduction of the number of packets transmitted at each space-batch (Figure \ref{Fig:HetTraffic_EB}). Regarding TP4, where the traffic load of each user is assigned randomly, the blocking probability is not distant from the one obtained with homogeneous traffic and, on average, is better than the performance with TP1, where four users have twice the traffic load of the rest. Moreover, it can be observed how the minimum blocking probability obtained using TP4 is the same as with homogeneous traffic, and that the maximum blocking probability  is only slightly higher than the one obtained using TP1.

%
%

\section{Conclusions}

We have presented a queuing model for the performance evaluation of SDMA-based Multiuser MPT systems using per-node FIFO packet scheduling. The analysis is built around a blind estimation of the space-batch size distribution, which is only based on the distribution of the traffic load between nodes, which is assumed to be homogeneous. This approximation allows to keep the model reasonably simple but, as the results show, also very accurate.

In such conditions, the model is expected to be used both alone, to evaluate the impact of the number of nodes, number of antennas, transmission rates, etc. on the system performance, or coupled with other link-layer mechanisms for the evaluation of more complex systems. For example, it can be easily combined with a model of the Distributed Coordination Function (DCF) \cite{zhao2011modeling,bellalta2013performance} to evaluate the performance in non-saturation conditions of the upcoming IEEE 802.11ac amendment \cite{IEEE80211ac}, that will support Multiuser MPT by using spatial multiplexing.

The presented model can be further extended in future works to cover other aspects, e.g.: 1) to consider non-uniform traffic distribution among nodes, as well as other Markov-based arrival processes, 2) to formulate schedulers that also consider {the existing} multiuser diversity (i.e., schedulers that pick the packets from the buffer based on the instantaneous CSI), 3) to combine the model with packet fragmentation and aggregation techniques in order to reduce the overheads due to CSI estimation and balance the duration of all transmissions using the individual transmission rates \cite{bellalta2012performance}, and 4) to consider different strategies to obtain and apply the CSI, including the case in which the AP is only able to use a set of pre-defined beamforming matrices, which affect the packets that can be selected for transmission, and the use of the Explicit Compressed Feedback protocol defined in the upcoming IEEE 802.11ac amendment.

%
%

\section*{Acknowledgements}

This work has been partially supported by the Spanish Government under projects TEC2012-32354 (Plan Nacional I+D), TEC2009-13000 (Plan Nacional I+D), CSD2008-00010 (Consolider-Ingenio Program) and by the Catalan Government (SGR2009\#00617).

%

\bibliographystyle{unsrt}
\bibliography{SMACMA}

\clearpage

\end{document}